# Twofold symmetry of *c*-axis resistivity in topological kagome superconductor $CsV_3Sb_5$ with in-plane rotating magnetic field

Ying Xiang[1†], Qing Li[1†], Yongkai Li[2,3†], Wei Xie[1], Huan Yang[1*], Zhiwei Wang[2,3*], Yugui Yao[2,3], and Hai-Hu Wen[1*]

[1]National Laboratory of Solid State Microstructures and Department of Physics, Collaborative Innovation Center of Advanced Microstructures, Nanjing University, Nanjing 210093, China

[2]Key Laboratory of Advanced Optoelectronic Quantum Architecture and Measurement, Ministry of Education, School of Physics, Beijing Institute of Technology, Beijing 100081, China

[3]Micronano Center, Beijing Key Lab of Nanophotonics and Ultrafine Optoelectronic Systems, Beijing Institute of Technology, Beijing 100081, China

[†]These authors contributed equally to this work. [*]e-mail: huanyang@nju.edu.cn; zhiweiwang@bit.edu.cn; hhwen@nju.edu.cn

**In transition metal compounds, due to the interplay of charge, spin, lattice and orbital degrees of freedom, many intertwined orders exist with close energies. One of the commonly observed states is the so-called nematic electron state, which breaks the in-plane rotational symmetry. This nematic state appears in cuprates, iron-based superconductor, etc. Nematicity may coexist, affect, cooperate or compete with other orders. Here we show the anisotropic in-plane electronic state and superconductivity in a recently discovered kagome metal $CsV_3Sb_5$ by measuring *c*-axis resistivity with the in-plane rotation of magnetic field. We observe a twofold symmetry of superconductivity in the superconducting state and a unique in-plane nematic electronic state in normal state when rotating the in-plane magnetic field. Interestingly these two orders are orthogonal to each other in terms of the field direction of the minimum resistivity. Our results shed new light in understanding non-trivial physical properties of $CsV_3Sb_5$.**

**Introduction**

Materials with a kagome lattice structure can host a rich variety of exotic states including spin liquid[1-3], spin density wave[4], charge density wave (CDW)[5,6], and superconductivity[4,5,7,8]. In addition, these kagome materials provide a novel platform to investigate topological electronic states[9-15]. Recently, a new family of kagome metals $A$V$_3$Sb$_5$ ($A$ = K, Rb, or Cs) has been discovered[16], and shortly afterwards, superconductivity is reported in this system[17-19]. The superconducting transition temperature ($T_c$) in CsV$_3$Sb$_5$ can be easily enhanced by pressure[20-23]. Although the superconductivity is argued to be a strong coupling one[24-26] probably with triplet pairing[27], the gap symmetry of this superconductor remains controversial[21,24-26,28-32] and requires further study. Besides the superconducting state, there is also a CDW transition in $A$V$_3$Sb$_5$[16-19]. The CDW order has been investigated experimentally[33-35] and theoretically[36,37]; it is probably driven by the nesting of saddle points near M points[18,35]. This issue and related electronic states of CDW are also investigated on atomic scale[24,25,38-40]. Furthermore, the existence of $Z_2$ topologically non-trivial states in $A$V$_3$Sb$_5$ has been evidenced by the observation of symmetry-protected Dirac crossing bands[17,19,41,42], giant anomalous Hall effect[43,44], a chiral charge order[38], and the possible existence of Majorana zero mode in the vortex core[25].

The nematic electronic state breaks the symmetry of the crystal structure in many strongly correlated electron systems[45], including cuprates[46,47], iron based superconductors[48,49], ultra-clean quantum Hall systems[50], Sr$_3$Ru$_2$O$_7$[51], etc. The twofold in-plane electronic anisotropy breaks the symmetry of the underlying lattice in these materials. Superconductivity with twofold symmetry seems to be a common feature in topological superconductors. Such feature is observed by different kinds of measurements in doped Bi$_2$Se$_3$[52-57] and heterostructures constructed by Bi$_2$Te$_3$ and high-temperature superconductors[58,59]. The nematic superconductivity is explained theoretically as a consequence of superconducting order parameter with odd parity derived from the spin-orbital coupling and the multi-orbital effect in these materials[60,61].

In this paper, by measuring the $c$-axis resistivity ($\rho_c$) using a Corbino-shape-like electrode configuration with in-plane rotating magnetic field, we observe a twofold rotational symmetry of $\rho_c$ both in the superconducting state and in the normal state of the topological kagome metal CsV$_3$Sb$_5$. The field direction for the minimum resistance in the superconducting state is along one pair of crystalline axes of the lattice (*a*-axis), which suggests a superconducting gap maximum in this direction. By applying a very strong in-plane magnetic field, we also observe a twofold rotational symmetry of $\rho_c$ in normal state but with an orthogonal direction of the minimum resistivity comparing to that in the superconducting state. The twofold rotational symmetry of $\rho_c$ may be related to the in-plane nematic electron state with the assistance of strong magnetic field. Furthermore, we find a six-fold oscillation component in the $\rho_c$ curve, which is due to the six-fold symmetry of the lattice. These findings contribute to better understanding of the electronic state and the superconductivity in this topological kagome metal.

**Results**

**Experimental configuration and superconducting characterization.** In order to detect the possible in-plane electronic anisotropy of the topological kagome metal CsV$_3$Sb$_5$, we measure the *c*-axis resistivity by using a Corbino-shape-like electrode configuration (Fig. 1a and Supplementary Fig. 1). The advantage of using this configuration is that a major part of current is flowing along *c*-axis, and thus is always perpendicular to the in-plane rotating magnetic field. This would avoid the undesired angle dependence of the in-plane resistivity due to the flux flow if the current were applied along *ab*-plane. Thus, the observed angle dependent variation of *c*-axis resistivity with in-plane rotating field can be safely attributed to some anisotropic electronic property in the material. Superconducting transition of the CsV$_3$Sb$_5$ single crystal is characterized by magnetization measurements (Fig. 1b), and the onset transition temperature is about 3.5 K determined from the enlarged view shown in the inset of Fig. 1b.

Figure 1c shows temperature dependence of in-plane ($\rho_{ab}$) and *c*-axis ($\rho_c$) resistivity. The normal-state resistivity shows a large anisotropy of $\alpha = \rho_c/\rho_{ab} = 23$ at 8

K, which suggests rather two-dimensionality of the material. The CDW transition can be clearly seen at about 95 K as an anomaly of resistivity, however, there is an obvious step-like increase in the $\rho_c(T)$ curve with decreasing temperature before the drop of resistivity. This feature is different from that measured by the in-plane resistivity which only exhibits a monotonic drop crossing the CDW transition. This difference has also been observed in the sister compound of RbV$_3$Sb$_5$[19]. Figure 1d,e show the temperature dependence of in-plane and $c$-axis resistivity, respectively; they are measured near the superconducting transition under different magnetic fields. The superconducting feature starts at about 3.5 K in $\rho_{ab}(T)$ at zero applied field, and the superconducting feature can be easily killed by a field of about 0.8 T at 2 K. All the results of $\rho_{ab}(T)$ are similar to those in previous reports[21,29]. For the $c$-axis resistivity, the detected zero-resistance temperature is the same as that measured with in-plane current. However, a superconducting-fluctuation-like behaviour can be seen obviously in $\rho_c(T)$ curves up to about 5.5 K. Furthermore, the resistivity drop at 2 K can be even seen under 7 T. This contrasting behaviour between $\rho_{ab}(T)$ and $\rho_c(T)$ near $T_c$ is quite interesting and deserves further study.

**Angular dependent of $c$-axis resistivity.** During the in-plane rotation of magnetic field, the initial field direction is set to be parallel to one of the sample edges (Fig. 1a), and it is found that this direction is just along one pair of in-plane crystallographic axes of the single crystal determined by Laue diffraction (see Supplementary Note 1). The angle dependent resistivity at 2 K (Fig. 2a) and different magnetic fields shows obvious twofold symmetry. At a field below 2.4 T, $\rho_c(\theta)$ curves show local minima near $\theta = 0º$ or 180º which is in the direction of one of the principal axes, namely $a$-axis. Since the resistivity minimum touches zero in the $\rho_c(\theta)$ curve measured at 0.4 T, the twofold symmetry of $\rho_c(\theta)$ curves is supposed to be induced by the anisotropic properties of the superconducting state. Here, on one particular curve of angle dependent resistivity, the minimum resistivity reflects a relatively larger upper critical field ($\mu_0 H_{c2}$). A simple consideration based on the Ginzburg-Landau theory and the Pippard definition $\xi \approx$ $\hbar v_\text{F}/\pi\Delta$ tells that $\mu_0 H_{c2} \propto \Delta^2/v_\text{F}^2$ with $\Delta$ the superconducting gap and $v_\text{F}$ the Fermi velocity[56]. Therefore, the gap maximum may be along *a*-axis, which suggests the possible existence of a twofold symmetry of superconductivity in $CsV_3Sb_5$.

In contrast, when the field exceeds 2.4 T, the orientation of field corresponding to the minimum resistivity become roughly orthogonal to that of superconducting state (below 2.4 T), i.e., $\rho_c(\theta)$ near the angles for the minimum of resistivity below 2.4 T now shows local maximum instead (Fig. 2a). This contrasting behaviour can be easily seen in the polar illustrations in Fig. 2b and 2c for the fields below and above 2.4 T, respectively. Although the $\rho_c(T)$ curve shows superconducting-fluctuation-like behaviour at fields stronger than 2.5 T at 2 K (Fig. 1d), the suppression of resistivity due to this effect is very weak when compared to the magnetoresistance. Thus the twofold symmetry of $\rho_c(\theta)$ curves at $\mu_0 H > 2.4$ T should be dominated by normal-state properties, which shows a clear two-fold symmetry. In addition, some extra oscillations can be seen in $\rho_c(\theta)$ curves measured at very high field, see for example Fig. 2a at 7 T. When we do Fourier transformation to the $\rho_c(\theta)$ curve measured at 7 T, the peak at 30º and 60º (Fig. 2d) suggest other symmetries. Actually the $\rho_c(\theta)$ oscillates by every 30º, which is strongly correlated with the arrangement of vanadium atoms in the lattice[17] (detailed analysis see Supplementary Note 2).

The two kinds of $\rho_c(\theta)$ curves below and above 2.4 T are roughly orthogonal to each other in terms of the field direction of the extremum resistivity. This tendency can be clearly seen from the field driven evolution of the angle dependent $\rho_c(\mu_0 H)$ curves (Fig. 3a). By using certain criterions, namely $\rho_c(\mu_0 H)$ = 100 and 5 μΩ·cm, we determined the angle dependent upper critical field $\mu_0 H_{c2}$ and zero-resistance field $\mu_0 H_0$ (Fig. 3b). Now the peak position corresponds to the minimum resistivity in the superconducting state. The in-plane anisotropy of $\mu_0 H_{c2}$ is about 1.2-1.3 which is consistent with the value determined from temperature dependent resistivity measurements (Supplementary Fig. 4d). The difference of minimum and maximum resistivity at these two typical angles ($\theta$ = 0º and 90º) and at 2 K is plotted in Fig. 3c with variation of field. One can see that there is a clear sign change at the field of about 2.4 T. However, when $T = 10$ K, we see no cross of $\rho_c(\mu_0 H)$ curves (Supplementary Fig. 4e). Now if we take the characteristic fields $\mu_0 H^*$ with the criterions of $\rho_c(\mu_0 H^*) = 300$ or 320 $\mu\Omega\cdot$cm, we see again the twofold feature of $\mu_0 H^*$ (Supplementary Fig. 4f), but now $\mu_0 H^*(\theta)$ shows oscillations with opposite phase as that of $\mu_0 H_{c2}(\theta)$.

**Temperature evolution of twofold feature.** Figure 4a and 4b show temperature evolution of $\rho_c(\theta)$ curves at two different fields of 0.4 T and 5 T, respectively. Obviously, the twofold feature of the $\rho_c(\theta)$ curve at 0.4 T weakens quickly with increase of temperature. When $T$ reaches about 4 K, this oscillation is greatly diminished. This indicates that the twofold symmetry of $\rho_c(\theta)$ is just induced by the flux flow dissipation in the superconducting state. However, at roughly the same angles for the minimum resistivity in the superconducting state, the resistivity peaks up when $\mu_0 H = 5$ T (Fig. 4b). Figure 4c shows the difference of $\rho_c(\theta)$ at the angles for minimum and maximum resistivity. Here the filled circles and squares represent the data for 0.4 T and 5 T, respectively. With increase of temperature, the twofold anisotropy of $\rho_c(\theta)$ curves measured at 0.4 T quickly vanishes around $T_c$. But that for 5 T changes much gently and finally disappears at about 60 K. This twofold symmetry at 5 T progressively weakens with temperature and may extend to the CDW transition temperature. In Fig. 5, we show the control experiment results carried out in another sample. One can also see the twofold symmetry of $\rho_c(\theta)$. The temperature evolution of the twofold feature is almost the same as the results shown in Fig. 4. All the observations are very similar in these two samples.

**Discussions**

Now we discuss the origin of the twofold symmetry observed in $\rho_c(\theta)$ curves. One may argue that this can be induced by the misalignment between the current direction and *c*-axis or that between the field and the *ab*-plane. Indeed, although we cannot avoid this misalignment, however, this is unlikely for our results because of the following reasons. Firstly, for the $\rho_c(\theta)$ curve measured at 7 T and 2 K, $\rho_c(\theta = 90º)/\rho_c(\theta$

= 0º) = 0.85 suggests a big difference. If the anisotropic magnetoresistance were induced by the misalignment between the magnetic field and the *ab*-plane, and the magnetoresistance were only induced by the *ab*-plane component of field which is perpendicular to the current, this anisotropy would correspond to a misalignment angle of about 32º (cos 32º = 0.85) between the field and the *ab*-plane. In the experiment, we can guarantee that the misalignment angle of the *ab*-plane to the field is less than 3°, thus it is impossible for such a big angle misalignment. Secondly, a relatively large normal-state resistance at 90° and 270° in the normal state would mean a large component of field perpendicular to the current (since we have a positive magnetoresistance), which would suppress superconductivity more severely and also induce a larger flux-flow resistivity at the same angle. But this contradicts the observations. Thirdly, we have repeated the experiments in the same sample (Supplementary Fig. 5) and another sample (Fig. 5) as a control experiment, and all show the same behaviours, indicating a high reproducibility. Although the directions of superconducting gap maximum are all along one pair of crystalline axes, these two directions have an intersect angle of about 60° (Supplementary Fig. 1) for the two samples which are mounted on the same sample holder. This excludes the possible error from the experimental setup.

It should be noted that the measurement is not about the in-plane resistivity of the sample, and the twofold symmetry in $\rho_c(\theta)$ curves may not reflect directly the nematic electron state in the *ab* plane. However, the twofold symmetry nature of $\rho_c(\theta)$ curves in the normal state only shows up when the magnetic field is stronger than 2.4 T. In the presence of magnetic field, the mobile electrons will possess a circular momentum in the plane perpendicular to the filed direction. Then the *c*-axis resistivity measured in this configuration should contain the contribution of the in-plane electronic states along the direction perpendicular to the magnetic field. For example, if the in-plane Fermi velocity has a two-fold symmetry, that will induce a twofold symmetric feature of the in-plane scattering rate, which can be detected by the *c*-axis resistivity with an in-plane magnetic field. Therefore, the twofold symmetry of $\rho_c(\theta)$ curves measured at high fields may suggest the in-plane nematic electronic state in the material. Since the twofold symmetry disappears at a temperature of about 60 K, the first scenario comes to our mind is that it may be related to the CDW state. We have been aware that there is an additional $4a_0$ unidirectional charge order besides the tri-directional charge order with a $2a_0$ period in this material[24]. This unidirectional charge order pattern corresponds to CDW stripes along *a*-axis, which may be explained based on the picture of topological CDW[62]. These CDW stripes induces a twofold symmetry of in-plane electronic properties, which behaves like a nematic state and has been detected by our *c*-axis resistivity when the magnetic field is rotated within the $V_3Sb_5$ planes. In addition, for the tri-directional charge order with a $2a_0$ period, the intensities of three sets of peaks have pronounced intensity anisotropy[24,38,40], which exactly shows the feature as the nematic CDW state. Besides, theoretically it was predicted that a chiral flux phase may exist and break the time-reversal symmetry in *c*-axis[63]. However, it should be noted that, in our experiment the magnetic field is applied in the *ab*-plane, how the in-plane magnetic field affects the chiral flux phase is unclear. Thus, an explicit theoretical picture of the spin origin for interpreting our observation is still lacking.

The twofold feature of $\rho_c(\theta)$ curve in the superconducting state may be intimately related to the feature in the normal state. A simple picture is that the CDW phase with twofold symmetry would gap out the density of states at the Fermi level leading to a truncated Fermi surface with twofold symmetry, this leads to a twofold symmetry of the superconducting gap or $H_{c2}$. By now the existence of the gap anisotropy is still hard to be detected directly from experiments[25]. Alternatively, concerning the fact that the nematic electronic state is observed with the help of a strong in-plane magnetic field with the absence of superconductivity, while the twofold symmetry of $H_{c2}(\theta)$ is observed in superconducting transition region with very small field: these two kinds of twofold symmetries may have different origins. We have noticed that the material is supposed or partly proved to be a topological superconductor[17,25,41,42,44], thus the superconductivity with twofold symmetry may be originated from the superconducting order parameter with odd parity. This scenario has recently been well proved[52-57,60,61] in topological superconductors $Cu_xBi_2Se_3$ or $Sr_xBi_2Se_3$. Beside these two possibilities, a $4a_0/3$ bidirectional pair density wave[24] or the spin-triplet superconductivity[27] may be extra reasons of the twofold symmetry of $\rho_c(\theta)$ curves at small magnetic field, thus quantitative analyses based on these models are highly desired. Our observations of twofold symmetry of superconductivity and the anti-phase oscillation of *c*-axis resistivity in normal state with respect to the in-plane magnetic field will shed new light in the study of this fascinating kagome and topological material.

**Methods**

**Single-crystal growth and preparation.** Single crystals of $CsV_3Sb_5$ were synthesized via a self-flux method[16]. The crystal orientation was determined by the Laue X-ray crystal alignment system (Photonic Science Ltd.). Some single crystals have naturally formed edges with the angle of about 120º for neighboured edges, and these edges are proved to be crystallographic axes by Laue diffraction measurements (see Supplementary Section 1).

**Resistivity measurements.** Resistance measurements were carried out in a physical property measurement system (PPMS, Quantum Design). Samples was cleaved along the Van der Waals layers, and some edge(s) were cut in order to form the hexagon structure by following the naturally formed edges (see Supplementary Fig. 1). The *c*-axis resistance was measured by the four-electrode method with the Corbino-shape-like configuration[56]. To eliminate the influence of the slight Hall signals on the raw data of angular dependence of resistivity, the resistivity taken at every angle has been averaged with positive and negative magnetic fields. The in-plane resistivity $\rho_{ab}$ was measured by the standard four-electrode method with remade electrodes on the same sample, and the current was in the *ab*-plane of the sample.

**Data availability**

Source data and all other data that support the plots within this paper and other findings of this study are available from the corresponding author upon reasonable request.

**Acknowledgements**

We acknowledge helpful discussions with Yaomin Dai. We appreciate the kind help in the analysis of the crystal alignment data given by Ning Yuan. This work was supported by National Key R&D Program of China (Nos. 2016YFA0300401, 2020YFA0308800), National Natural Science Foundation of China (Nos. 11927809, 11974171, 12061131001, 92065109, 11734003 and 1904294), Strategic Priority Research Program (B) of Chinese Academy of Sciences (No. XDB25000000), Beijing Natural Science Foundation (No. Z190006), and Beijing Institute of Technology Research Fund Program for Young Scholars (No. 3180012222011).

**Author contributions**

Y.L., Z.W, and Y.Y. grew samples. Q.L. and W.X. measured and analyzed the crystal orientation. Y.X., H.Y., and H.-H.W. carried out resistivity measurements. H.Y., H.-H.W., Y.X., and Q.L. analyzed the data and wrote the manuscript which was proof-read and agreed by all authors.

**Competing interests**

The authors declare no competing interests.

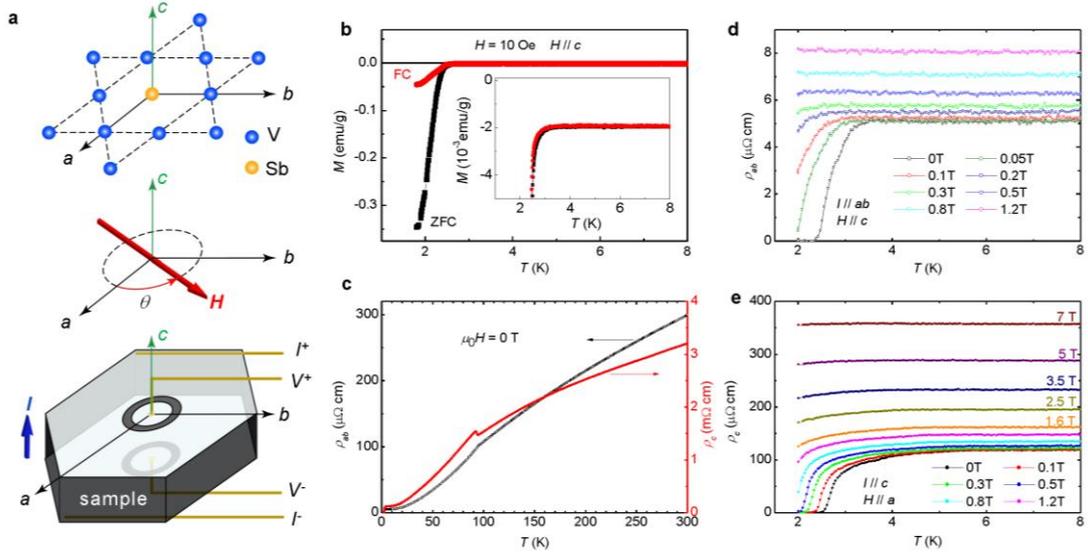

**Fig. 1 | Measurement configuration of *c*-axis resistivity and characterization on superconductivity. a**, The *c*-axis resistivity of CsV$_3$Sb$_5$ is measured by using a Corbino-shape-like electrode configuration. The electric current is applied mainly along *c*-axis of the single crystal, and the magnetic field is applied parallel to and rotated in the *ab*-plane. Single crystals usually have naturally formed edges with the angle of about 120º for neighboured edges, and these edges are along directions of crystallographic axes (see Supplementary Section 1). **b**, Temperature dependent magnetization measured with the zero-field-cooling (ZFC) and the field-cooling (FC) modes. **c**, Temperature dependence of *c*-axis and in-plane resistivity which are measured with different configurations (Supplementary Fig. 1). **d,e**, Temperature dependence of in-plane (**d**) and *c*-axis (**e**) resistivity measured near the superconducting transition temperature at different magnetic fields.

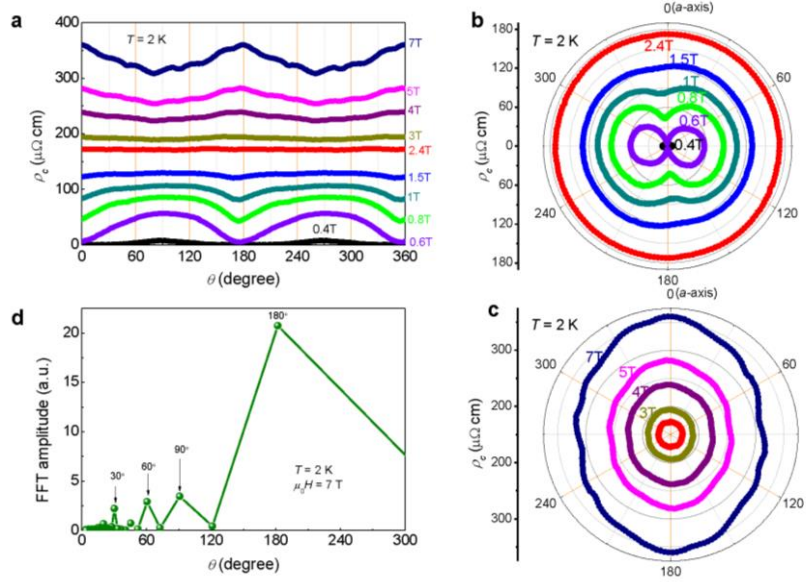

**Fig. 2 | Twofold symmetry of angular dependent *c*-axis resistivity under in-plane magnetic field. a**, Angular dependence of *c*-axis resistivity measured at different in-plane magnetic fields. The featureless $\rho_c(\theta)$ curve measured at 2.4 T separates two sets of $\rho_c(\theta)$ curves holding phase-reversed oscillations with twofold symmetry. **b,c**, Angular dependent *c*-axis resistivity plotted in polar coordinate measured with rotating in-plane magnetic field of the magnitude (**b**) below and (**c**) above 2.4 T. Local minima in curves measured at $\mu_0 H < 2.4$ T change to local maxima in curves at $\mu_0 H > 2.4$ T in the direction along *a*-axis. In addition, $\rho_c(\theta)$ curves measured at $\mu_0 H > 2.4$ T show extra oscillation of six-fold symmetry besides the major twofold signal. **d**, Fourier transformation result to the $\rho_c(\theta)$ curve measured at 7 T. The period of 60° suggests a six-fold symmetry.

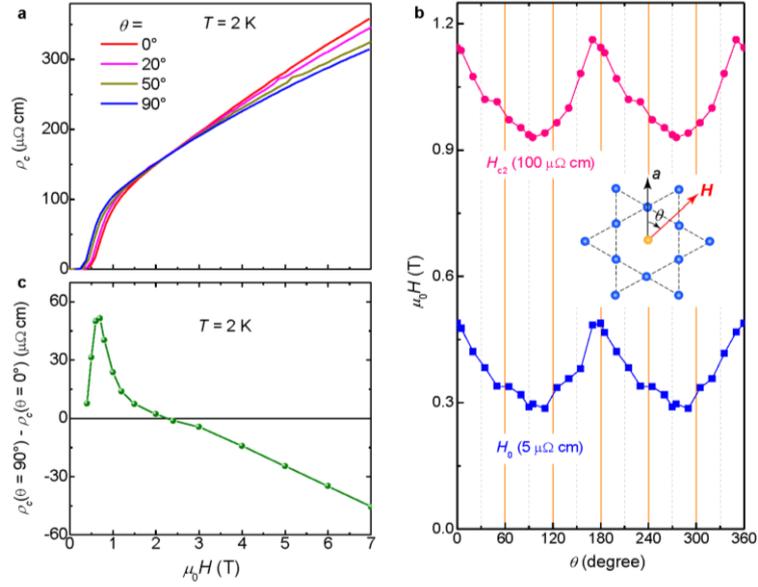

**Fig. 3 | Magnetic-field induced phase reverse of $\rho_c(\theta)$ curves with twofold symmetry. a**, Magnetic field dependence of *c*-axis resistivity measured with different angles between field and the *a*-axis at $T = 2$ K. **b**, Angular dependent upper critical field and zero-resistance field ($\mu_0H_0$) by using different criteria of *c*-axis resistivity. **c**, Field dependence of the *c*-axis-resistivity difference between $\theta = 0º$ and $90º$. The resistivity difference changes its sign at a field of about 2.4 T.

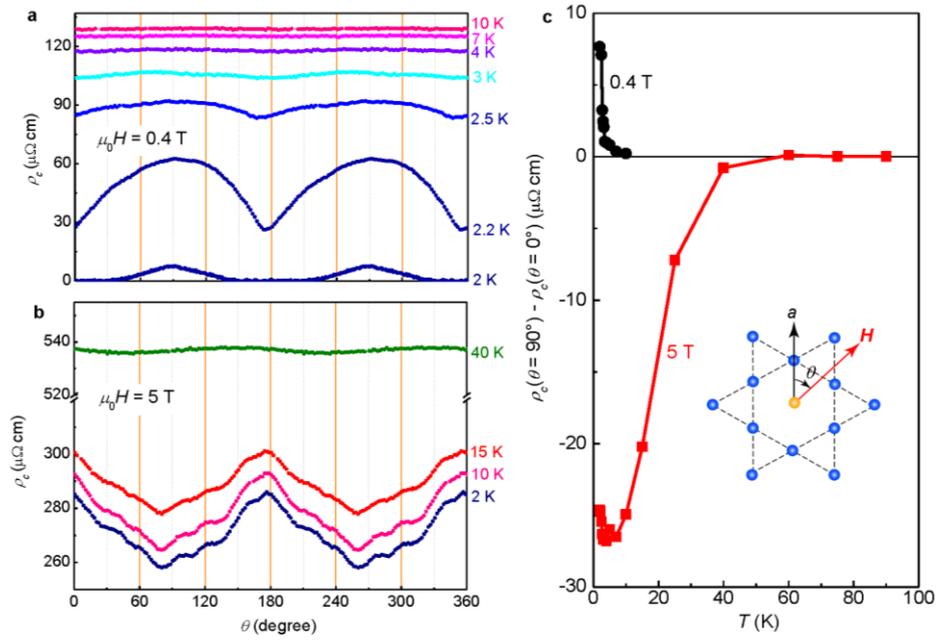

**Fig. 4 | Temperature evolution of the nematicity of *c*-axis resistivity. a,b,** Angular dependent *c*-axis resistivity measured at different temperatures under a magnetic field of (**a**) 0.4 and (**b**) 5 T. **c**, Temperature dependence of nematicity of *c*-axis resistivity between $\theta = 0°$ and $90°$.

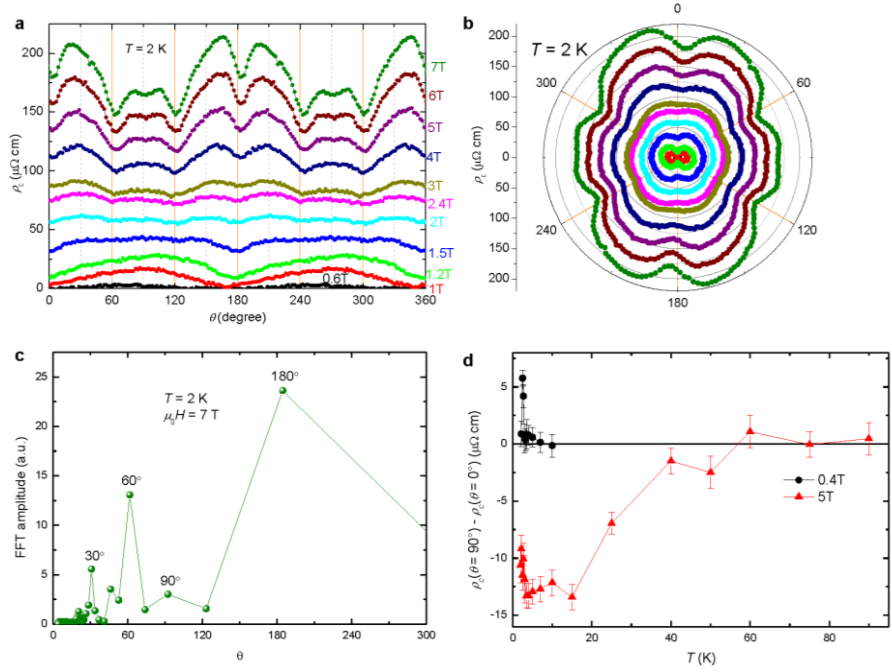

**Fig. 5 | Control experiment of $\rho_c(\theta)$ curves with twofold symmetry of in another sample. a,b**, Angular dependence of *c*-axis resistivity measured at different in-plane magnetic field shown by a rectangular and a polar coordinate, respectively. The inverse of the local extrema can be observed when the magnetic field crosses about 2 T. **c**, Fourier transformation result to the $\rho_c(\theta)$ curve measured at 7 T. **d**, Temperature dependence of nematicity of *c*-axis resistivity between $\theta = 0º$ and $90º$. The error bars in the figure are determined by the noise for the resistivity measurement. The relatively small thickness and large surface area make the resistance very small with a relatively large noise.